\documentclass[11pt]{article}

\usepackage[T1]{fontenc}
\usepackage{lmodern}
\usepackage[margin=1in]{geometry}
\usepackage{amsmath,amssymb,mathtools}
\usepackage{microtype}
\usepackage{booktabs}
\usepackage{graphicx}
\usepackage{tikz}
\usepackage{xcolor}
\usepackage{cite}
\usepackage[hypertexnames=false]{hyperref}
\usepackage[nameinlink,capitalise,noabbrev]{cleveref}
\usepackage{titlesec}
\usepackage{float}
\usepackage{array}

\numberwithin{equation}{section}
\allowdisplaybreaks

\newcommand{\cN}{\mathcal N}
\newcommand{\dd}{\mathrm d}
\newcommand{\ii}{\mathrm i}
\DeclareMathOperator{\Res}{Res}
\DeclareMathOperator{\erfc}{erfc}

\DeclareMathOperator{\He}{He}
\newcommand{\maybeincludegraphics}[2][]{%
  \IfFileExists{#2}{\includegraphics[#1]{#2}}{%
    \fbox{\parbox[c][0.24\textheight][c]{0.8\textwidth}{\centering Missing figure file\\[0.4em]\texttt{\detokenize{#2}}}}%
  }%
}

\hypersetup{
  colorlinks=true,
  linkcolor=blue,
  citecolor=blue,
  urlcolor=blue,
  pdftitle={The double-logarithmic four-graviton Regge sector as a rank-two twisted period system},
  pdfauthor={Agustin Sabio Vera},
  pdfsubject={Regge limit, supergravity, double logarithms, twisted periods, Mellin transform, intersection theory},
  pdfkeywords={Regge limit, supergravity, double logarithms, Mellin transform, twisted periods, intersection theory, Weber equation, symbols},
  pdfcreator={LaTeX}
}

\crefname{equation}{Eq.}{Eqs.}
\Crefname{equation}{Eq.}{Eqs.}

\titleformat{\section}{\large\normalfont\bfseries}{\thesection}{1em}{}
\titleformat{\subsection}{\normalsize\normalfont\bfseries}{\thesubsection}{1em}{}
\titleformat{\subsubsection}{\normalsize\normalfont\bfseries}{\thesubsubsection}{1em}{}

\makeatletter
\renewcommand{\maketitle}{
\begin{center}
{\LARGE\normalfont \@title \par}
\vspace{1em}
{\large\normalfont \@author \par}
\vspace{1em}
{\normalsize\normalfont \@date \par}
\vspace{2em}
\end{center}
}
\makeatother

\title{\bf The double-logarithmic four-graviton\\ Regge sector as a rank-two twisted period system}
\author{Agust\'\i n Sabio Vera\\
{\small Instituto de F\'\i sica Te\'orica UAM/CSIC, Nicol\'as Cabrera 13-15, E-28049 Madrid, Spain}\\
{\small Theoretical Physics Department, Universidad Aut\'onoma de Madrid, E-28049 Madrid, Spain}\\
{\small \texttt{agustin.sabio@uam.es}}}
\date{\today}

\begin{document}
\maketitle

\begin{center}
{\normalsize\normalfont Abstract\par}
\vspace{0.8em}
\begin{minipage}{0.88\textwidth}
We study the double-logarithmic four-graviton Regge sector in \(\cN\)-extended supergravity. Its Mellin-space solution is already known in terms of parabolic-cylinder functions. We show that the same answer can be organized as a rank-two twisted period system, meaning that two closely related weighted integrals determine the full Mellin partial wave. These functions satisfy a simple pair of first-order differential equations and a recursion as the number of supersymmetries \(\cN\) changes. This gives a uniform description of the full supergravity family, clarifies the relation between the positive-ray Euler integral and the earlier contour representation, and reproduces the same reduction rule through intersection theory. The reformulation also makes the special cases with four and six supersymmetries particularly transparent and yields a simple Hermite-polynomial construction for the low-even theories.
\end{minipage}
\end{center}

\section{Introduction}

Gravity at high energy has been studied for a long time. Early work on graviton reggeization, multi-Regge processes and related effective descriptions goes back to Grisaru, van Nieuwenhuizen and Wu, to Lipatov, and to later developments of the corresponding high-energy framework \cite{Grisaru1975tb,Lipatov1982vv,Lipatov1982it,Lipatov1991nf,Lipatov2011ab,Lipatov2016nvu}. Early supergravity studies also pointed to nontrivial bound-state and Regge phenomena \cite{Grisaru1981ra}. More broadly, the infrared and exponentiation properties of gravitational amplitudes belong to a line of work that starts with Weinberg's classic analysis of soft gravitons and continues into modern eikonal and exponentiation studies \cite{Weinberg1965,DiVecchia2019myk,DiVecchia2019kta,DiVecchiaHeissenbergRussoVeneziano2024}. 

Modern amplitude methods have greatly expanded this subject. String-inspired and loop-based techniques have played a central role in gravity amplitudes for decades \cite{Bern1991an,Bern1993wt,Dunbar1994bn,Bern2009kd,Abreu2020lyk}. Exact integrated four-graviton amplitudes at two and three loops provide especially sharp checks of any all-order proposal \cite{BoucherVeronneauDixon2011,HennMistlberger2019}. Related advances include graviton emission in the Regge limit, color-kinematics and double-copy constructions, and soft-graviton double-copy structures \cite{GravitonEmission2012,CKRegge2013,Johansson2013nsa,Johansson2013aca,SabioVeraVazquezMozo2015,BernDaviesDennenHuang2012}. The same broader landscape also includes Wilson-line approaches to high-energy gravity, multi-Regge analyses beyond four points, all-loop relations between gauge theory and supergravity in the Regge limit, and recent shockwave, effective-theory, and high-energy-structure developments \cite{MelvilleNaculichSchnitzerWhite2014,CaronHuot2020vlo,Naculich2020clm,Raj2023irr,Raj2024xsi,Bondarenko2024lzx,Rothstein2024nlq,Ciafaloni:2018uwe,Barcaro2025ifi,Alessio2025isu,Alessio2026bdi}.

Within this context, the double-logarithmic sector is worth studying in its own right. The reason is that it is one of the few places where an all-order resummation problem can be posed precisely and then checked against explicit loop amplitudes. The double-logarithmic mechanism itself was clarified long ago in gauge theory by Kirschner and Lipatov \cite{Kirschner1982xw,Kirschner1983di}. In gravity, the Regge limit means that one keeps the momentum transfer \(t<0\) fixed and sends \(s\) to infinity. In that limit, the terms with two powers of \(\ln(s/(-t))\) at each loop order form a closed subsector. The present paper studies that subsector for four-graviton scattering in \(\cN\)-extended supergravity.

For the four-graviton amplitude, the Mellin-space evolution equation in Einstein gravity and in extended supergravity was derived in \cite{BartelsLipatovSabioVera2014}. The original analysis already reduced the Mellin problem to Riccati form and then to a second-order linear equation, wrote the physical solution through parabolic-cylinder functions, and gave a contour-integral representation valid for arbitrary \(\cN\) \cite{BartelsLipatovSabioVera2014}. Later work emphasized the same pole structure and its physical consequences, especially for the \(\cN\ge 4\) theories \cite{SabioVera2019,SabioVera2020}. Here we do not claim to find a new solution. The goal is different. We want to identify the simplest organizing structure behind that known answer and to explain why the same Mellin problem can be described by only two closely related functions.

The central observation is that the full solution is controlled by two neighboring weighted integrals. We refer to these integrals as periods. The adjective twisted only means that the integrand carries a nontrivial weight. In the present problem that weight is the factor \(z^{\eta-1}e^{-z^2/2-\sigma z}\). The phrase rank two means that only two neighboring period functions are needed to reconstruct the full answer. Their ratio gives the Mellin partial wave, which is the basic function entering the Mellin representation of the amplitude. Those same two functions satisfy a first-order differential system in the variable \(\sigma\), and they are also linked by a recursion when the number of supersymmetries \(\cN\) changes. This is why the period language is useful. It turns one known special-function solution into a framework that treats the whole supergravity family in a uniform way.

It is useful to mention explicitly what was already known and what is new here. Known from previous work are the Mellin evolution equation, its Riccati reduction, the parabolic-cylinder solution, the contour representation, and the special nodes \(\cN=4\) and \(\cN=6\). New here are the normalized-period formulation as the primary object, the fact that the whole problem closes on two neighboring periods, the differential system that they satisfy, the recursion that connects neighboring values of \(\cN\), the precise comparison between the positive-ray integral and the older contour representation, the independent derivation of the same reduction through twisted intersection theory, and the resulting Hermite-polynomial construction of the low-even theories.

This point of view naturally requires analytic continuation in the theory label. Physical theories correspond to integer values of \(\cN\), but the normalized periods introduced below define an analytic family in the parameter \(\cN\). The unnormalized integral representatives are then meromorphic because gamma functions appear in the normalization. This analytically continued family is the natural setting for the differential system, for the recursion in \(\cN\), and for the discussion of special values such as \(\cN=4\) and \(\cN=6\).

The mathematical framework used below comes from a branch of analysis and geometry designed to study weighted integrals \cite{AomotoKita2011,ChoMatsumoto2016,Matsumoto1998}. In that language, a hypergeometric integral is simply an integral whose integrand contains nontrivial powers or exponentials of the integration variable. Twisted cohomology is a way to organize which differential forms give the same weighted integral after integration by parts. Rapid-decay cycles are integration contours chosen so that the exponential factor makes the integral converge at infinity. Intersection pairings measure how such objects overlap and, in practice, allow one to reduce more complicated integrals to a small basis.

In modern amplitude theory this language became useful when it was realized that these intersection pairings can be used to project Feynman integrals onto a chosen basis and to derive differential equations for them \cite{Mizera2018intersection,MastroliaMizera2019,Frellesvig2019maxcut,Frellesvig2019vectorspace,Frellesvig2021multivariate,Mizera2020status}. For the present paper, the key point is straightforward. The same reduction identity that closes the Mellin problem can also be recovered from this second method, which is more geometric in spirit. This matters because it gives an independent check of the reduction and turns it into a systematic procedure. For regulated Gaussian twists and their intersection-theoretic treatment in quantum mechanics and quantum field theory, see also \cite{CacciatoriMastrolia2022}. The present graviton problem gives a closely related one-variable example in which a regulated Gaussian-type weight again leads to a finite-dimensional twisted cohomology and to a compact reduction to a small basis of periods.

Because the integrand in our problem decays along the positive real axis, one also needs the language of rapid-decay homology and irregular connections, which is the framework adapted to integrals with exponential falloff at infinity \cite{BlochEsnault2004,Hien2007}. In addition, the even ladder passes through resonant values. These are special parameter values where the simplest contour formulas may degenerate or need to be regularized. For that reason, regularizable cycles and relative twisted homology also enter in the background \cite{MimachiYoshida2007,Matsumoto2018relative}. Standard special-function facts used later, such as asymptotics and the location of zeros, are taken from the DLMF \cite{DLMF}.

The four-graviton channel \(A_4(1^-,2^-,3^+,4^+)\) is used throughout for comparison with exact amplitudes. \Cref{sec:mellin} recalls the Mellin equation, derives the exact recursion for the perturbative coefficients, records the fixed-loop dictionary used to compare the resummed result with the explicit two-loop and three-loop amplitudes of \cite{BoucherVeronneauDixon2011,HennMistlberger2019}, and ends with a short well-known symbol-level cross-check in maximal supersymmetry using the symbol technology of \cite{Duhr2012}. \Cref{sec:twistedperiods} then performs the central reformulation. It promotes the normalized periods to the primary objects, relates the positive-ray integral and the contour representation, and derives the first-order differential system satisfied by the two neighboring periods. \Cref{sec:analytic} develops the recursion in \(\cN\), the special nodes \(\cN=4\) and \(\cN=6\), the Mellin poles and residues, and the explicit construction of the low-even sector. \Cref{sec:intersection} rederives the same reduction through intersection theory and includes the local Laurent expansions needed for that computation.

The paper can therefore be read in two complementary ways. From the physics side, it gives a more transparent description of a known exact double-logarithmic sector and organizes its special theories and pole structure more cleanly. From the mathematical side, it provides a compact worked example in which special functions, weighted integrals, differential systems, recursions in the number of supersymmetries, rapid-decay cycles, and intersection theory all meet in one variable. The claim is not that the four-graviton double-logarithmic sector was unsolved. The claim is that its known solution has a simpler and more revealing organization than had previously been made explicit.

\section{Mellin equation, exact recursion and fixed-loop checks}
\label{sec:mellin}

\subsection*{Kinematics and Mellin representation}

We begin with the standard Regge kinematics for four-graviton scattering in four dimensions. The Mandelstam invariants satisfy \(s+t+u=0\). The Regge limit keeps \(t<0\) fixed and sends \(s\) to infinity. It is convenient to introduce
\begin{equation}
 x\equiv -\frac{t}{s},
 \qquad
 x\to 0^+
 \qquad \text{in the physical Regge region.}
\end{equation}
For later comparison with explicit amplitudes we use the MHV component \(\mathcal A_4(1^-,2^-,3^+,4^+)\) \cite{BoucherVeronneauDixon2011,HennMistlberger2019}. Following \cite{BartelsLipatovSabioVera2014,SabioVera2019,SabioVera2020}, we factor the full amplitude into a Born term and a dimensionless correction,
\begin{equation}
 \mathcal A_4^{(\cN)}(s,t)=\mathcal A_4^{\mathrm{Born}}(s,t)\,M_4^{(\cN)}(s,t),
 \qquad
 \mathcal A_4^{\mathrm{Born}}(s,t)=\kappa^2\frac{s^3}{tu},
\end{equation}
with \(\kappa^2=8\pi G\). The parameters used throughout are
\begin{equation}
 \alpha=\frac{\kappa^2}{8\pi^2},
 \qquad
 b\equiv -\alpha t>0,
 \qquad
 L\equiv \ln\!\left(\frac{s}{-t}\right),
 \qquad
 g\equiv \alpha t\,L^2,
 \qquad
 \xi\equiv \frac{\alpha t}{2}\,L^2=\frac{g}{2}.
 \label{eq:alphab}
\end{equation}

The infrared-finite double-logarithmic factor admits the Mellin representation
\begin{equation}
 M_{4,\mathrm{DL}}^{(\cN)}(s,t)=
 \int_{\delta-\ii\infty}^{\delta+\ii\infty}\frac{\dd\omega}{2\pi\ii}
 \left(\frac{s}{-t}\right)^\omega \frac{f_\omega^{(\cN)}}{\omega},
 \qquad
 \delta>0,
 \label{eq:mellinrep}
\end{equation}
as derived in \cite{BartelsLipatovSabioVera2014}. This form is useful because inverse powers of \(\omega\) turn directly into powers of the large logarithm. Indeed,
\begin{equation}
 \int_{\delta-\ii\infty}^{\delta+\ii\infty}\frac{\dd\omega}{2\pi\ii}
 \left(\frac{s}{-t}\right)^\omega \frac{1}{\omega^{2n+1}}
 =\frac{1}{(2n)!}\ln^{2n}\!\left(\frac{s}{-t}\right),
 \qquad n\ge 0.
 \label{eq:basicmellin}
\end{equation}
It is worth stating explicitly the physical meaning of the Mellin variable \(\omega\). If we write
\begin{equation}
 Y\equiv \ln\!\left(\frac{s}{-t}\right),
 \qquad
 \left(\frac{s}{-t}\right)^\omega=e^{\omega Y},
\end{equation}
then \(\omega\) is the Mellin-conjugate variable to the large Regge logarithm \(Y\). In that sense, singularities in the complex \(\omega\)-plane control the asymptotic high-energy behavior, while inverse powers of \(\omega\) generate powers of \(Y\). Near the graviton exchange point, one may think of \(\omega\) as the shift of the complex angular momentum away from \(j=2\), namely \(j=2+\omega\). The Mellin poles discussed below are therefore the Regge singularities of this double-logarithmic sector \cite{BartelsLipatovSabioVera2014}.
This makes it natural to expand \(f_\omega^{(\cN)}\) as
\begin{equation}
 f_\omega^{(\cN)}=\sum_{n=0}^{\infty} C_n^{(\cN)}
 \left(\frac{\alpha t}{\omega^2}\right)^n,
 \qquad
 C_0^{(\cN)}=1.
 \label{eq:fexpansion}
\end{equation}
The coefficients \(C_n^{(\cN)}\) are the all-order data of the double-logarithmic sector. Substituting \cref{eq:fexpansion} into \cref{eq:mellinrep} gives
\begin{equation}
 M_{4,\mathrm{DL}}^{(\cN)}(s,t)
 =1+\sum_{n\ge 1}\frac{C_n^{(\cN)}}{(2n)!}
 \left(\alpha t\,\ln^2\!\frac{s}{-t}\right)^n.
 \label{eq:mdlseries}
\end{equation}
The problem is therefore reduced to finding these coefficients exactly.

\subsection*{Exact recursion and special cases}

The Mellin-space evolution equation found in \cite{BartelsLipatovSabioVera2014} is
\begin{equation}
 f_\omega^{(\cN)}
 =1+\alpha t\left[
 \eta\,\frac{\bigl(f_\omega^{(\cN)}\bigr)^2}{\omega^2}
 -\frac{\dd}{\dd\omega}\left(\frac{f_\omega^{(\cN)}}{\omega}\right)
 \right],
 \qquad
 \eta=\frac{\cN-6}{2}.
 \label{eq:evolution}
\end{equation}
All dependence on the theory enters through the single number \(\eta\). In particular,
\begin{equation}
 \cN=8 \Longleftrightarrow \eta=1,
 \qquad
 \cN=6 \Longleftrightarrow \eta=0,
 \qquad
 \cN=4 \Longleftrightarrow \eta=-1.
\end{equation}
Substituting the series \cref{eq:fexpansion} into \cref{eq:evolution} and matching equal powers of \(\alpha t/\omega^2\) gives the exact recursion
\begin{equation}
 C_m^{(\cN)}
 =(2m-1)C_{m-1}^{(\cN)}
 +\eta\sum_{k=0}^{m-1}C_k^{(\cN)}C_{m-1-k}^{(\cN)},
 \qquad
 m\ge 1.
 \label{eq:recursion}
\end{equation}
The first terms are
\begin{equation}
 C_1^{(\cN)} = 1+\eta,
 \qquad
 C_2^{(\cN)} = (1+\eta)(3+2\eta),
 \qquad
 C_3^{(\cN)} = (1+\eta)(5\eta^2+17\eta+15).
 \label{eq:Csvarias}
\end{equation}
For maximal supersymmetry this gives \(C_1^{(8)}=2\), \(C_2^{(8)}=10\), and \(C_3^{(8)}=74\).

The recursion already shows two important special cases. When \(\cN=6\), one has \(\eta=0\), so the quadratic term disappears and the recursion becomes
\begin{equation}
 C_m^{(6)}=(2m-1)C_{m-1}^{(6)},
 \qquad
 C_0^{(6)}=1,
 \label{eq:N6recurrence}
\end{equation}
hence
\begin{equation}
 C_m^{(6)}=(2m-1)!!.
 \label{eq:N6doublefact}
\end{equation}
The theory with \(\cN=4\) is also special. Its exact cancellation will be recovered later from the closed-form solution, but the first coefficients in \cref{eq:Csvarias} already show the pattern that leads to it.

\subsection*{The known Riccati solution and the present focus}

It is helpful to recall the form of the known solution before reinterpreting it. The original analysis introduced a rescaled Mellin variable and showed that the evolution equation can be rewritten as a Riccati equation, meaning a first-order nonlinear differential equation with a quadratic term. That equation can then be linearized, which leads to Weber's equation and to the parabolic-cylinder functions that solve it \cite{BartelsLipatovSabioVera2014}. The same solution was later used to study the pole structure of the \(\cN\ge 4\) theories \cite{SabioVera2019,SabioVera2020}.

This paper starts where that story stops. We do not look for a different special function. Instead, we ask whether the known answer is already the visible part of a simpler framework. The claim developed below is that the Weber solution can be written in terms of two neighboring weighted integrals, that these two functions satisfy a closed first-order differential system, and that the same reduction also generates a recursion in the number of supersymmetries.

\subsection*{Loop coefficients and fixed-loop dictionary}

For comparison with explicit amplitudes it is useful to translate the recursion coefficients into the language used in fixed-loop calculations. The coefficient of the highest power \(\ln^{2L}(s/(-t))\) at loop order \(L\) is \(C_L^{(\cN)}(\alpha t)^L/(2L)!\). In practice, comparisons are cleaner in the logarithm of the resummed factor. We write
\begin{equation}
 \ln M_{4,\mathrm{DL}}^{(\cN)}(s,t)
 =\sum_{n\ge 1}r_n^{(\cN)}g^n
 =\sum_{n\ge 1}\widehat r_n^{(\cN)}\xi^n,
 \qquad
 \widehat r_n^{(\cN)}=2^n r_n^{(\cN)}.
 \label{eq:rdef}
\end{equation}
Expanding \(\ln M\) through cubic order gives
\begin{equation}
 \widehat r_1^{(\cN)}=C_1^{(\cN)},
 \qquad
 \widehat r_2^{(\cN)}=\frac{C_2^{(\cN)}}{6}-\frac{(C_1^{(\cN)})^2}{2},
 \qquad
 \widehat r_3^{(\cN)}=\frac{C_3^{(\cN)}}{90}-\frac{C_1^{(\cN)}C_2^{(\cN)}}{6}+\frac{(C_1^{(\cN)})^3}{3}.
 \label{eq:rhatdefs}
\end{equation}
For \(\cN=8\) one finds
\begin{equation}
 \ln M_{4,\mathrm{DL}}^{(8)}(s,t)
 =2\xi-\frac{1}{3}\xi^2+\frac{7}{45}\xi^3+\mathcal O(\xi^4)
 =g-\frac{1}{12}g^2+\frac{7}{360}g^3+\mathcal O(g^4).
 \label{eq:N8logseries}
\end{equation}
The corresponding fixed-loop coefficients are collected in \cref{tab:fixedloop}. The exact two-loop amplitudes of Boucher-Veronneau and Dixon reproduce the entries for \(\cN=4,5,6,8\) \cite{BoucherVeronneauDixon2011}. For maximal supersymmetry the three-loop amplitude of Henn and Mistlberger reproduces the coefficient \(\widehat r_3^{(8)}=7/45\), or equivalently \(r_3^{(8)}=7/360\) in the variable \(g\) \cite{HennMistlberger2019}.

\begin{table}[H]
\centering
\small
\renewcommand{\arraystretch}{1.2}
\begin{tabular}{@{}>{\centering\arraybackslash}m{0.06\textwidth}>{\raggedright\arraybackslash}m{0.33\textwidth}>{\raggedright\arraybackslash}m{0.29\textwidth}>{\raggedright\arraybackslash}m{0.18\textwidth}@{}}
\toprule
\(\cN\) &
\(M_{4,\mathrm{DL}}^{(\cN)}(s,t)\) through \(\xi^3\) &
\(\ln M_{4,\mathrm{DL}}^{(\cN)}(s,t)\) through \(\xi^3\) &
loop-order check \\
\midrule
\(4\) &
\(1\) &
\(0\) &
exact cancellation \\
\(5\) &
\(1+\frac{1}{2}\xi+\frac{1}{6}\xi^2+\frac{31}{720}\xi^3\) &
\(\frac{1}{2}\xi+\frac{1}{24}\xi^2+\frac{1}{720}\xi^3\) &
two loops \\
\(6\) &
\(1+\xi+\frac{1}{2}\xi^2+\frac{1}{6}\xi^3\) &
\(\xi\) &
two loops \\
\(8\) &
\(1+2\xi+\frac{5}{3}\xi^2+\frac{37}{45}\xi^3\) &
\(2\xi-\frac{1}{3}\xi^2+\frac{7}{45}\xi^3\) &
two and three loops \\
\bottomrule
\end{tabular}
\caption{Double-logarithmic coefficients in the normalization commonly used for comparison with exact integrated amplitudes. For \(\cN=4\), the entries \(M_{4,\mathrm{DL}}^{(4)}=1\) and \(\ln M_{4,\mathrm{DL}}^{(4)}=0\) reflect the exact cancellation of the whole double-logarithmic sector.}
\label{tab:fixedloop}
\end{table}

\subsection*{Symbol-level cross-check in maximal supersymmetry}

The fixed-loop checks can also be written in the language of symbols \cite{Duhr2012}. This is useful because the symbol immediately isolates the highest power of the Regge logarithm. Let \(x=-t/s\). Since \(\mathcal S(\ln^n x)=n!\,x^{\otimes n}\), the leading double logarithm at \(L\) loops is the coefficient of the repeated word \(x^{\otimes 2L}\).

At two loops Boucher-Veronneau and Dixon write the \(\cN=8\) remainder in terms of three pure weight-four functions whose symbols are \cite{BoucherVeronneauDixon2011}
\begin{equation}
 \mathcal S(f_1)=x\otimes x\otimes x\otimes \frac{x}{1-x},
 \label{eq:Sf1}
\end{equation}
\begin{equation}
 \mathcal S(f_2)=x\otimes x\otimes x\otimes (1-x),
 \label{eq:Sf2}
\end{equation}
\begin{equation}
 \mathcal S(f_3)= -
 \frac{x}{1-x}\otimes \frac{x}{1-x}\otimes \frac{x}{1-x}\otimes (1-x).
 \label{eq:Sf3}
\end{equation}
Their remainder may be written as
\begin{equation}
 F_{4,\cN=8}^{(2)}(s,t,u)
 =
 8\Bigl(
 t u\,[f_1(x)+f_1(1-x)]
 +s u\,[f_2(x)+f_3(x)]
 +s t\,[f_2(1-x)+f_3(1-x)]
 \Bigr).
 \label{eq:F2symbolform}
\end{equation}
In the Regge limit only the repeated \(x\)-letters contribute to the leading logarithm. This leaves
\begin{equation}
 \mathcal S\!\left(F_{4,\cN=8}^{(2)}\right)\Big|_{x^{\otimes 4}}
 =8\bigl(tu+st\bigr)x^{\otimes 4}
 =-8t^2x^{\otimes 4},
 \label{eq:F2symbolprojection}
\end{equation}
and therefore
\begin{equation}
 F_{4,\cN=8}^{(2)}
 =-\frac{t^2}{3}\ln^4\!\left(\frac{s}{-t}\right)
 +\mathcal O\!\left(\ln^3\!\frac{s}{-t}\right).
 \label{eq:F2leadinglog}
\end{equation}
This agrees with \(\widehat r_2^{(8)}=-1/3\).

At three loops Henn and Mistlberger organize the finite remainder through a last-entry representation \cite{HennMistlberger2019}
\begin{equation}
 f^{(3)}(x_{\mathrm{HM}})
 =
 C^{(3)}+
 \int_0^{x_{\mathrm{HM}}} \frac{\dd x'}{x'}\,g^{(3)}(x'),
 \label{eq:HMf3integral}
\end{equation}
so that
\begin{equation}
 \mathcal S\!\left(f^{(3)}\right)
 =
 \mathcal S\!\left(g^{(3)}\right)\otimes x_{\mathrm{HM}}.
 \label{eq:HMlastentrysymbol}
\end{equation}
The leading Regge logarithm is therefore the coefficient of \(x_{\mathrm{HM}}^{\otimes 6}\). Their Regge expansion gives
\begin{equation}
 F_4^{(3)}(s,t,u)
 =
 \frac{7}{360}\,t^3 \log^6\!\left(\frac{s}{-t}\right)
 +\mathcal O\!\left(\ln^5\!\frac{s}{-t}\right),
 \label{eq:HM3loopleadingours}
\end{equation}
which is exactly the three-loop coefficient encoded in \cref{eq:N8logseries}.

\section{From Mellin to Twisted Periods}
\label{sec:twistedperiods}

\subsection*{The Weber reduction}

We begin by recalling the standard route from the Mellin equation to the known parabolic-cylinder solution. Nothing in this subsection is new. Its purpose is to fix notation and to prepare the reformulation in terms of weighted integrals.

It is convenient to divide the Mellin partial wave by \(\omega\) and define
\begin{equation}
 F(\omega)\equiv \frac{f_\omega^{(\cN)}}{\omega}.
 \label{eq:Fdef}
\end{equation}
In terms of \(F(\omega)\), the Mellin equation \cref{eq:evolution} takes the form
\begin{equation}
 \frac{\dd F}{\dd\omega}
 =\eta F^2+\frac{\omega F-1}{b}.
 \label{eq:Fode}
\end{equation}
This already shows the basic structure of the problem. The equation is first order, but it is nonlinear. The natural dimensionless Mellin variable is
\begin{equation}
 \sigma=\frac{\omega}{\sqrt b}
 =\frac{\omega}{\sqrt{-\alpha t}}.
 \label{eq:sigmadef}
\end{equation}
For \(\eta\neq 0\), it is also convenient to absorb the overall normalization by defining
\begin{equation}
 R(\sigma)
 =\eta\sqrt b\,F(\omega)
 =\eta\sqrt{-\alpha t}\,\frac{f_\omega^{(\cN)}}{\omega}.
 \label{eq:Rdef}
\end{equation}
Then \cref{eq:Fode} becomes
\begin{equation}
 \frac{\dd R}{\dd\sigma}
 =R^2+\sigma R-\eta.
 \label{eq:riccati}
\end{equation}
This is the Riccati equation already identified in \cite{BartelsLipatovSabioVera2014}. The physical solution is selected by the original Mellin representation. One requires \(f_\omega^{(\cN)} \to 1\) for \(|\omega|\to\infty\) and \(\Re(\omega)>0\), which implies
\begin{equation}
 F(\omega)\sim \frac{1}{\omega},
 \qquad
 R(\sigma)\sim \frac{\eta}{\sigma}
 \qquad \text{for} \qquad
 |\sigma|\to\infty,
 \quad \Re(\sigma)>0.
 \label{eq:Rbc}
\end{equation}

The next step is standard. A Riccati equation can be turned into a linear second-order equation by writing the unknown function as a logarithmic derivative. We set
\begin{equation}
 R(\sigma)=-\frac{\dd}{\dd\sigma}\ln I_0(\sigma),
\end{equation}
hence
\begin{equation}
 R'(\sigma)=-\frac{I_0''(\sigma)}{I_0(\sigma)}+R(\sigma)^2.
\end{equation}
Substituting this into \cref{eq:riccati} gives
\begin{equation}
 I_0''(\sigma)-\sigma I_0'(\sigma)-\eta I_0(\sigma)=0.
 \label{eq:linearI0}
\end{equation}
At this point the nonlinear problem has been replaced by a linear one.

It is now useful to remove a simple Gaussian factor. Write
\begin{equation}
 I_0(\sigma,\eta)=e^{\sigma^2/4}\psi(\sigma,\eta).
 \label{eq:gaugetransform}
\end{equation}
A short calculation shows that the first-derivative term then disappears, and \cref{eq:linearI0} becomes
\begin{equation}
 \psi''(\sigma,\eta)
 +\left(\frac{1}{2}-\eta-\frac{\sigma^2}{4}\right)\psi(\sigma,\eta)
 =0.
 \label{eq:Weber}
\end{equation}
This is Weber's equation. Its solutions are the parabolic-cylinder functions that appeared in the original analysis.

So far we have only recovered the known special-function form of the answer. The reason for recalling these steps is that they also show where the later integral representation comes from. The linear equation \cref{eq:linearI0} is the natural starting point for introducing weighted integrals, and the Weber equation \cref{eq:Weber} tells us which special-function family those integrals must reproduce. The next subsection explains that the same solution can be organized by two neighboring weighted integrals, and that those two objects already determine the full Mellin partial wave.

\subsection*{Normalized periods and contour integrals}

We now rewrite the solution of Weber’s equation in a form that makes the rank-two structure explicit. Let \(D_\nu(\sigma)\) denote the standard parabolic-cylinder function solving \cref{eq:Weber}. Restoring the Gaussian factor introduced in Eq.~(3.10), we therefore define the normalized periods
\begin{equation}
\widehat{J}_N(\sigma)\equiv e^{\sigma^2/4}D_{(6-N)/2}(\sigma).
\label{eq:Jhatdef}
\end{equation}
The neighboring pair \(\widehat{J}_N(\sigma)\) and \(\widehat{J}_{N+2}(\sigma)\) will be the two basic period functions that determine the full Mellin partial wave. This normalization is convenient because the gamma-function prefactors appear only in the convergent integral representation written below, while \(\widehat{J}_N(\sigma)\) itself varies analytically with \(N\).

The Mellin partial wave is then
\begin{equation}
 \frac{f_\omega^{(\cN)}}{\omega}
 =
 \frac{1}{\sqrt{-\alpha t}}\,
 \frac{\widehat J_{\cN+2}(\sigma)}{\widehat J_{\cN}(\sigma)}.
 \label{eq:fJhatratio}
\end{equation}
This is still the known Weber solution. The point is that it is now written as the ratio of two neighboring period functions. To recover an integral representation, start from the linear equation \cref{eq:linearI0} and look for a Laplace-type solution
\begin{equation}
 I_0(\sigma)=\int_\gamma e^{-\sigma z}K(z)\,\dd z,
 \label{eq:laplaceansatz}
\end{equation}
with a contour \(\gamma\) on which boundary terms vanish. This ansatz is natural because differentiation with respect to \(\sigma\) inserts powers of \(z\), which is exactly what will generate the neighboring moments. One finds
\begin{equation}
 I_0'(\sigma)=-\int_\gamma z\,e^{-\sigma z}K(z)\,\dd z,
 \qquad
 I_0''(\sigma)=\int_\gamma z^2 e^{-\sigma z}K(z)\,\dd z.
\end{equation}
Using one integration by parts in \cref{eq:linearI0}, one sees that \(K(z)\) must satisfy
\begin{equation}
 zK'(z)+(z^2+1-\eta)K(z)=0.
\end{equation}
Its solution is
\begin{equation}
 K(z)=C\,z^{\eta-1}e^{-z^2/2}.
\end{equation}
Hence the natural weight is
\begin{equation}
 u(z,\sigma,\eta)=z^{\eta-1}\exp\!\left(-\frac{z^2}{2}-\sigma z\right).
 \label{eq:twist}
\end{equation}
This is the origin of the weighted integrals that appear throughout the paper.

The first two moments of this weight are
\begin{equation}
 I_0(\sigma,\eta)
 =\int_0^\infty z^{\eta-1}e^{-z^2/2-\sigma z}\,\dd z,
 \label{eq:I0def}
\end{equation}
\begin{equation}
 I_1(\sigma,\eta)
 =\int_0^\infty z^{\eta}e^{-z^2/2-\sigma z}\,\dd z.
 \label{eq:I1def}
\end{equation}
At the physical value \(\eta=\eta_{\cN}=(\cN-6)/2\), they become
\begin{equation}
 J_{\cN}(\sigma)=I_0(\sigma,\eta_{\cN}),
 \qquad
 J_{\cN+2}(\sigma)=I_1(\sigma,\eta_{\cN}).
 \label{eq:momentcontiguity}
\end{equation}
For \(\Re(\eta_{\cN})>0\) and \(\Re(\sigma)>0\), these integrals converge absolutely on the positive real axis, and one has
\begin{equation}
 J_{\cN}(\sigma)\equiv \Gamma(\eta_{\cN})\,\widehat J_{\cN}(\sigma)=I_0(\sigma,\eta_{\cN}),
 \label{eq:Jndef}
\end{equation}
\begin{equation}
 J_{\cN+2}(\sigma)=\Gamma(\eta_{\cN}+1)\,\widehat J_{\cN+2}(\sigma)=I_1(\sigma,\eta_{\cN}).
 \label{eq:Jnp2def}
\end{equation}
Equivalently,
\begin{equation}
 J_{\cN}(\sigma)=\Gamma(\eta_{\cN})e^{\sigma^2/4}D_{(6-\cN)/2}(\sigma),
 \qquad
 J_{\cN+2}(\sigma)=\Gamma(\eta_{\cN}+1)e^{\sigma^2/4}D_{(4-\cN)/2}(\sigma).
 \label{eq:parabolic}
\end{equation}
The positive real axis is therefore the most direct integral representation of the two basic periods.

It is also useful to compare this explicit positive-ray formula with the contour representation used in the original Mellin analysis \cite{BartelsLipatovSabioVera2014}. Choose the branch
\begin{equation}
 z^{\eta-1}=\exp \bigl((\eta-1)\log \,z\bigr),
 \qquad
 0<\arg z<2\pi,
\end{equation}
so that the branch cut lies on \(\mathbb R_+\). Let \(\mathcal C\) be the contour which comes from \(+\infty\) to \(0\) just below the cut, circles the origin clockwise, and returns from \(0\) to \(+\infty\) just above the cut. The phase jump across the cut comes entirely from the algebraic factor \(z^{\eta-1}\). The exponential factor \(e^{-z^2/2-\sigma z}\) is single-valued. Writing the contour as the sum of its lower-bank piece, its small circle around the origin, and its upper-bank piece, one finds
\begin{align}
 \int_{\mathcal C} z^{\eta-1}e^{-z^2/2-\sigma z}\,\dd z
 &=
 \int_{\infty}^{\varepsilon}
 e^{2\pi\ii\eta}x^{\eta-1}e^{-x^2/2-\sigma x}\,\dd x
 +
 \int_{|z|=\varepsilon}
 z^{\eta-1}e^{-z^2/2-\sigma z}\,\dd z
 \notag\\
 &\hspace{2.7em}
 +
 \int_{\varepsilon}^{\infty}
 x^{\eta-1}e^{-x^2/2-\sigma x}\,\dd x.
 \label{eq:contourEulerRelationExpanded}
\end{align}
For \(\Re(\eta)>0\), the small-circle term vanishes as \(\varepsilon\to 0\), because near the origin the integrand behaves like \(z^{\eta-1}\dd z\). The factor \(1-e^{2\pi\ii\eta}\) is the difference between the upper-bank and lower-bank contributions once the lower bank picks up the monodromy phase. One then obtains
\begin{equation}
 \int_{\mathcal C} z^{\eta-1}e^{-z^2/2-\sigma z}\,\dd z
 =
 \bigl(1-e^{2\pi\ii\eta}\bigr)
 \int_0^\infty z^{\eta-1}e^{-z^2/2-\sigma z}\,\dd z,
 \qquad
 \Re(\eta)>0,\ \Re(\sigma)>0.
 \label{eq:contourEulerRelation}
\end{equation}
Away from resonant values of \(\eta\), this gives
\begin{equation}
 \widehat J_{\cN}(\sigma)
 =
 \frac{1}{\Gamma(\eta_{\cN})\bigl(1-e^{2\pi\ii\eta_{\cN}}\bigr)}
 \int_{\mathcal C} z^{\eta_{\cN}-1}e^{-z^2/2-\sigma z}\,\dd z.
 \label{eq:JhatContour}
\end{equation}
The positive-ray integral and the older contour formula are two faces of the same analytically continued period. The one on the positive real axis is the simplest when the integral converges directly. The contour \(\mathcal C\) keeps track of the monodromy around the branch point and continues the same object beyond that naive convergence region.

\subsection*{One-step reduction and the rank-two differential system}

A key fact is the reduction of the third moment to the first two. Define
\begin{equation}
 I_2(\sigma,\eta)=\int_0^\infty z^{\eta+1}e^{-z^2/2-\sigma z}\,\dd z.
 \label{eq:I2def}
\end{equation}
Then
\begin{equation}
 I_0''(\sigma,\eta)
 =-\frac{\dd}{\dd\sigma}I_1(\sigma,\eta)
 =I_2(\sigma,\eta).
 \label{eq:I0ppI2}
\end{equation}
At the physical value of \(\eta\), this is simply \(J_{\cN+4}(\sigma)=I_2(\sigma,\eta_{\cN})\). Now consider the total derivative
\begin{equation}
 \frac{\dd}{\dd z}\Bigl(z^{\eta_{\cN}}e^{-z^2/2-\sigma z}\Bigr)
 =z^{\eta_{\cN}}
 \left(\eta_{\cN} z^{-1} -z-\sigma \right)e^{-z^2/2-\sigma z}.
 \label{eq:totalderivative}
\end{equation}
Integrating it over the positive real axis gives
\begin{equation}
 I_2(\sigma,\eta_{\cN})=\eta_{\cN} I_0(\sigma,\eta_{\cN})-\sigma I_1(\sigma,\eta_{\cN}).
 \label{eq:I2reduction}
\end{equation}
This becomes
\begin{equation}
 J_{\cN+4}(\sigma)=\frac{\cN-6}{2}\,J_{\cN}(\sigma)-\sigma J_{\cN+2}(\sigma).
 \label{eq:Jrecurrence}
\end{equation}
This is the fundamental one-step reduction. Differentiating \cref{eq:Jndef,eq:Jnp2def} and using \cref{eq:Jrecurrence}, we obtain
\begin{equation}
 \frac{\dd}{\dd\sigma}
 \begin{pmatrix}
 J_{\cN}\\[0.3em]
 J_{\cN+2}
 \end{pmatrix}
 =
 \begin{pmatrix}
 0 & -1\\[0.3em]
 \dfrac{6-\cN}{2} & \sigma
 \end{pmatrix}
 \begin{pmatrix}
 J_{\cN}\\[0.3em]
 J_{\cN+2}
 \end{pmatrix}.
 \label{eq:PfaffianRaw}
\end{equation}
This first-order system is what is usually called a Gauss--Manin system\footnote{The two functions \(J_{\cN}(\sigma)\) and \(J_{\cN+2}(\sigma)\) form a rank-two family of periods depending on the parameter \(\sigma\), and \cref{eq:PfaffianRaw} is the connection that describes how this basis varies with \(\sigma\). In that sense it is a Gauss--Manin connection. For the general construction of Gauss--Manin connections in families, see Katz and Oda~\cite{KatzOda1968}. For period integrals of the same general hypergeometric type as those considered here, see Aomoto~\cite{Aomoto1987GaussManin}.}. In the present setting, this just means that it describes how the two basic weighted integrals vary with \(\sigma\). After dividing by gamma factors, one gets
\begin{equation}
 \frac{\dd}{\dd\sigma}
 \begin{pmatrix}
 \widehat J_{\cN}\\[0.3em]
 \widehat J_{\cN+2}
 \end{pmatrix}
 =
 \begin{pmatrix}
 0 & \dfrac{6-\cN}{2}\\[0.3em]
 -1 & \sigma
 \end{pmatrix}
 \begin{pmatrix}
 \widehat J_{\cN}\\[0.3em]
 \widehat J_{\cN+2}
 \end{pmatrix}.
 \label{eq:PfaffianJhat}
\end{equation}
The phrase rank two now has a concrete meaning. Once these two neighboring functions are known, the whole solution follows from them.

Taking the ratio of the two entries of \cref{eq:PfaffianRaw} reproduces the Riccati equation. In the same way,
\begin{equation}
 \frac{J_{\cN+2}(\sigma)}{J_{\cN}(\sigma)}
 =
 \eta_{\cN}\,
 \frac{\widehat J_{\cN+2}(\sigma)}{\widehat J_{\cN}(\sigma)},
\end{equation}
so \cref{eq:fJhatratio} is equivalent to
\begin{equation}
 \frac{f_\omega^{(\cN)}}{\omega}
 =
 \frac{1}{\eta_{\cN}\sqrt{-\alpha t}}\,
 \frac{J_{\cN+2}(\sigma)}{J_{\cN}(\sigma)}
 =
 -\frac{1}{\eta_{\cN}\sqrt{-\alpha t}}\,
 \frac{\dd}{\dd\sigma}\ln J_{\cN}(\sigma),
 \qquad \eta_{\cN}\neq 0.
 \label{eq:fperiodratio}
\end{equation}
The case \(\eta_{\cN}=0\) will be handled separately in \cref{subsec:specialtheories}.

\subsection*{Selection of the physical branch}

The positive real axis is convenient but it also selects the physical branch. Using \cref{eq:parabolic} and the standard asymptotic form of the parabolic-cylinder function,
\begin{equation}
 D_\nu(\sigma)\sim e^{-\sigma^2/4}\sigma^\nu
 \left(1+\mathcal O(\sigma^{-2})\right),
 \qquad
 |\sigma|\to\infty,
 \qquad
 |\arg \sigma|<\frac{3\pi}{4},
\end{equation}
see \S 12.9 of the DLMF \cite{DLMF}, one finds
\begin{equation}
 J_{\cN}(\sigma)
 =
 \Gamma(\eta_{\cN})\sigma^{-\eta_{\cN}}
 \left(1+\mathcal O(\sigma^{-2})\right),
 \label{eq:I0asymptotic}
\end{equation}
and
\begin{equation}
 J_{\cN+2}(\sigma)
 =
 \Gamma(\eta_{\cN}+1)\sigma^{-\eta_{\cN}-1}
 \left(1+\mathcal O(\sigma^{-2})\right).
 \label{eq:I1asymptotic}
\end{equation}
Therefore
\begin{equation}
 \frac{J_{\cN+2}(\sigma)}{J_{\cN}(\sigma)}
 =\frac{\eta_{\cN}}{\sigma}
 \left(1+\mathcal O(\sigma^{-2})\right),
 \label{eq:Rasymptoticperiod}
\end{equation}
which agrees with the physical behavior in \cref{eq:Rbc}. This is also why one has a rapid-decay contour. The factor \(e^{-z^2/2-\sigma z}\) decays exponentially along the positive real axis when \(\Re(\sigma)>0\). In the language of Bloch--Esnault and Hien, the positive real axis is therefore a contour along which the weighted integrand dies off fast enough to make the integral well behaved \cite{BlochEsnault2004,Hien2007}.

\section{Discrete contiguity and resonant nodes}
\label{sec:analytic}

\subsection*{Discrete contiguity in \texorpdfstring{$\cN$}{N}}

The one-step reduction does more than give a differential system in \(\sigma\). It also gives a recursion that moves from one theory to the next. This is what is usually called a contiguity relation. In the present context it is the statement that neighboring values of \(\cN\) are linked by a second-order recursion. Removing the gamma factors from \cref{eq:Jrecurrence} gives
\begin{equation}
 \widehat J_{\cN}(\sigma)
 -\sigma \widehat J_{\cN+2}(\sigma)
 +\frac{4-\cN}{2}\,\widehat J_{\cN+4}(\sigma)
 =0.
 \label{eq:Jhatrecurrence}
\end{equation}
This is the usual three-term recursion of the parabolic-cylinder function rewritten in the theory label \(\cN\). Taking ratios turns it into a discrete nonlinear map,
\begin{equation}
 \frac{\widehat J_{\cN+4}(\sigma)}{\widehat J_{\cN+2}(\sigma)}
 =
 \frac{2}{\cN-4}
 \left(
 \frac{\widehat J_{\cN}(\sigma)}{\widehat J_{\cN+2}(\sigma)}
 -\sigma
 \right).
 \label{eq:discreteRiccatiJhat}
\end{equation}
away from zeros of the denominators. This is the discrete analogue of the Riccati equation in \(\sigma\).

Two values of \(\cN\) stand out. The first is \(\cN=6\), where
\begin{equation}
 \widehat J_6(\sigma)=1.
 \label{eq:J6base}
\end{equation}
This is the simplest even theory and serves as the natural starting point of the even recursion. The second is \(\cN=4\), where the coefficient of \(\widehat J_{\cN+4}\) vanishes. The recursion then reduces to
\begin{equation}
 \widehat J_4(\sigma)=\sigma \widehat J_6(\sigma)=\sigma.
 \label{eq:J4sigmaJ6}
\end{equation}
Using \cref{eq:fJhatratio} one obtains
\begin{equation}
 \frac{f_\omega^{(4)}}{\omega}
 =
 \frac{1}{\sqrt{-\alpha t}}\,
 \frac{\widehat J_6(\sigma)}{\widehat J_4(\sigma)}
 =
 \frac{1}{\omega},
\end{equation}
hence
\begin{equation}
 f_\omega^{(4)}\equiv 1.
 \label{eq:N4exactcancellation}
\end{equation}
So the exact cancellation of the double-logarithmic sector in \(\cN=4\) is built directly into the discrete recursion.

The same recursion generates the even theories below \(\cN=6\) one step at a time,
\begin{equation}
 \widehat J_6(\sigma)=1,
 \qquad
 \widehat J_4(\sigma)=\sigma,
 \qquad
 \widehat J_2(\sigma)=\sigma^2-1,
 \qquad
 \widehat J_0(\sigma)=\sigma^3-3\sigma.
 \label{eq:evenladder}
\end{equation}
These are precisely the first probabilists' Hermite polynomials. More generally, for \(m\ge 0\),
\begin{equation}
 \widehat J_{6-2m}(\sigma)=\He_m(\sigma).
 \label{eq:Heidentification}
\end{equation}
This identification also gives a direct partial-fraction expansion for the Mellin partial wave. For \(m\ge 1\), write
\begin{equation}
 \He_m(\sigma)=\prod_{j=1}^{m}\bigl(\sigma-\sigma_j^{(m)}\bigr),
 \qquad
 \He_m'(\sigma)=m\,\He_{m-1}(\sigma),
 \label{eq:Hemfactorization}
\end{equation}
where \(\sigma_j^{(m)}\) are the \(m\) simple zeros of \(\He_m\). Then
\begin{equation}
 \frac{f_\omega^{(6-2m)}}{\omega}
 =
 \frac{1}{\sqrt{-\alpha t}}\,
 \frac{\He_{m-1}(\sigma)}{\He_m(\sigma)}
 =
 \frac{1}{m\sqrt{-\alpha t}}\,
 \frac{\dd}{\dd\sigma}\ln \He_m(\sigma)
 =
 \frac{1}{m}
 \sum_{j=1}^{m}
 \frac{1}{\omega-\sqrt{-\alpha t}\,\sigma_j^{(m)}}.
 \label{eq:evenHermitePartialFraction}
\end{equation}
Thus every low-even theory is described by \(m\) simple Mellin poles located at
\begin{equation}
 \omega_j=\sqrt{-\alpha t}\,\sigma_j^{(m)},
 \qquad j=1,\dots,m,
 \label{eq:omegajdef}
\end{equation}
with pole positions given by the zeros of the corresponding Hermite polynomial.

The weight \(z^{\eta-1}\) also records how the family behaves when one goes once around the origin. The resulting factor is
\begin{equation}
 M_0=e^{2\pi\ii\eta_{\cN}}.
 \label{eq:localmonodromyN}
\end{equation}
For integer \(\cN\) this becomes \(M_0=(-1)^{\cN}\). Hence the physical family splits into an even ladder and an odd ladder,
\begin{equation}
 \cN \text{ even}: \qquad M_0=+1,
 \qquad
 \cN \text{ odd}: \qquad M_0=-1.
 \label{eq:parityladders}
\end{equation}
The exponential factor at infinity is the same in every theory. What changes with \(\cN\) is the power of \(z\) near the origin. That is why the theories are close to each other and yet not identical.

Two different issues must be kept separate. Outside \(\Re(\eta)>0\), the positive-ray integral \(\int_0^\infty u(z,\sigma,\eta)\,\dd z\) no longer converges absolutely and one must use analytic continuation in \(\eta\). Independently, on the even ladder the monodromy factor equals one. This is the setting in which regularizable cycles and relative twisted homology enter the background \cite{MimachiYoshida2007,Matsumoto2018relative}. For the present paper, the practical point is simpler. The normalized periods \(\widehat J_{\cN}\) remain the global objects, while the positive-ray integral is a convenient local representation whenever it converges.

\subsection*{Mellin poles, asymptotics and the \texorpdfstring{$\cN=4$}{N=4} threshold}

The Mellin poles are governed by the zeros of the denominator in the period ratio. Since \(\widehat J_{\cN}(\sigma)=e^{\sigma^2/4}D_{(6-\cN)/2}(\sigma)\), they are the zeros of a single parabolic-cylinder function. These zeros are simple. If both the function and its derivative vanished at the same point, the uniqueness theorem for Weber's equation would force the whole solution to vanish identically.

From \cref{eq:fperiodratio}, near a simple zero \(\sigma_k\) of \(J_{\cN}\),
\begin{equation}
 J_{\cN}(\sigma)=J_{\cN}'(\sigma_k)(\sigma-\sigma_k)+\mathcal O\bigl((\sigma-\sigma_k)^2\bigr),
\end{equation}
and since \(J_{\cN+2}=-J_{\cN}'\), one gets
\begin{equation}
 \frac{J_{\cN+2}(\sigma)}{J_{\cN}(\sigma)}
 =-\frac{1}{\sigma-\sigma_k}+\mathcal O(1).
 \label{eq:universalpole}
\end{equation}
If \(\omega_k=\sigma_k\sqrt{-\alpha t}\), then
\begin{equation}
 \Res_{\omega=\omega_k}\left(\frac{f_\omega^{(\cN)}}{\omega}\right)
 =-\frac{1}{\eta}.
 \label{eq:universalresidue}
\end{equation}
So the residues at simple Mellin poles are universal. They do not depend on the position of the pole.

The large-\(|\sigma|\) behavior also follows directly from the Riccati equation. Look for an odd inverse-power series,
\begin{equation}
 R(\sigma)=\frac{a_1}{\sigma}+\frac{a_3}{\sigma^3}+\frac{a_5}{\sigma^5}
 +\mathcal O(\sigma^{-7}),
 \qquad
 |\sigma|\to\infty,
 \quad \Re(\sigma)>0.
 \label{eq:Ransatz}
\end{equation}
Substituting into \cref{eq:riccati} gives
\begin{equation}
 a_1=\eta,
 \qquad
 a_3=-\eta(\eta+1),
 \qquad
 a_5=\eta(\eta+1)(2\eta+3),
\end{equation}
hence
\begin{equation}
 R(\sigma)
 =\frac{\eta}{\sigma}
 -\frac{\eta(\eta+1)}{\sigma^3}
 +\frac{\eta(\eta+1)(2\eta+3)}{\sigma^5}
 +\mathcal O(\sigma^{-7}).
 \label{eq:Rasymptoticseries}
\end{equation}
Using \(\sigma=\omega/\sqrt{-\alpha t}\), this becomes
\begin{equation}
 \frac{f_\omega^{(\cN)}}{\omega}
 =
 \frac{1}{\omega}
 +(\eta+1)\frac{\alpha t}{\omega^3}
 +(\eta+1)(2\eta+3)\frac{(\alpha t)^2}{\omega^5}
 +\mathcal O(\omega^{-7}),
 \label{eq:matchrecursionfromasymptotics}
\end{equation}
which matches the perturbative recursion.

A particularly important change takes place at \(\cN=4\). It is convenient to write
\begin{equation}
 \nu\equiv -\eta=\frac{6-\cN}{2},
\end{equation}
so that the denominator is \(D_\nu(\sigma)\). The DLMF classification of real zeros gives
\begin{equation}
 \nu>1 \Longrightarrow D_\nu \text{ has positive real zeros},
 \qquad
 \nu=1 \Longrightarrow D_1(0)=0,
\end{equation}
\begin{equation}
 0<\nu<1 \Longrightarrow D_\nu \text{ has no positive real zeros},
 \qquad
 \nu\le 0 \Longrightarrow D_\nu \text{ has no real zeros}.
 \label{eq:DLMFthreshold}
\end{equation}
Translated back to \(\cN\), this becomes
\begin{equation}
 \cN<4 \Longrightarrow \text{positive real zeros},
 \qquad
 \cN=4 \Longrightarrow \sigma=0 \text{ threshold zero},
\end{equation}
\begin{equation}
 4<\cN<6 \Longrightarrow \text{no positive real zeros},
 \qquad
 \cN\ge 6 \Longrightarrow \text{no real zeros}.
 \label{eq:Nthresholdclassification}
\end{equation}
This threshold is important because it separates theories with positive-real Mellin poles from theories without them.

\begin{figure}[H]
\centering
\maybeincludegraphics[width=0.94\textwidth]{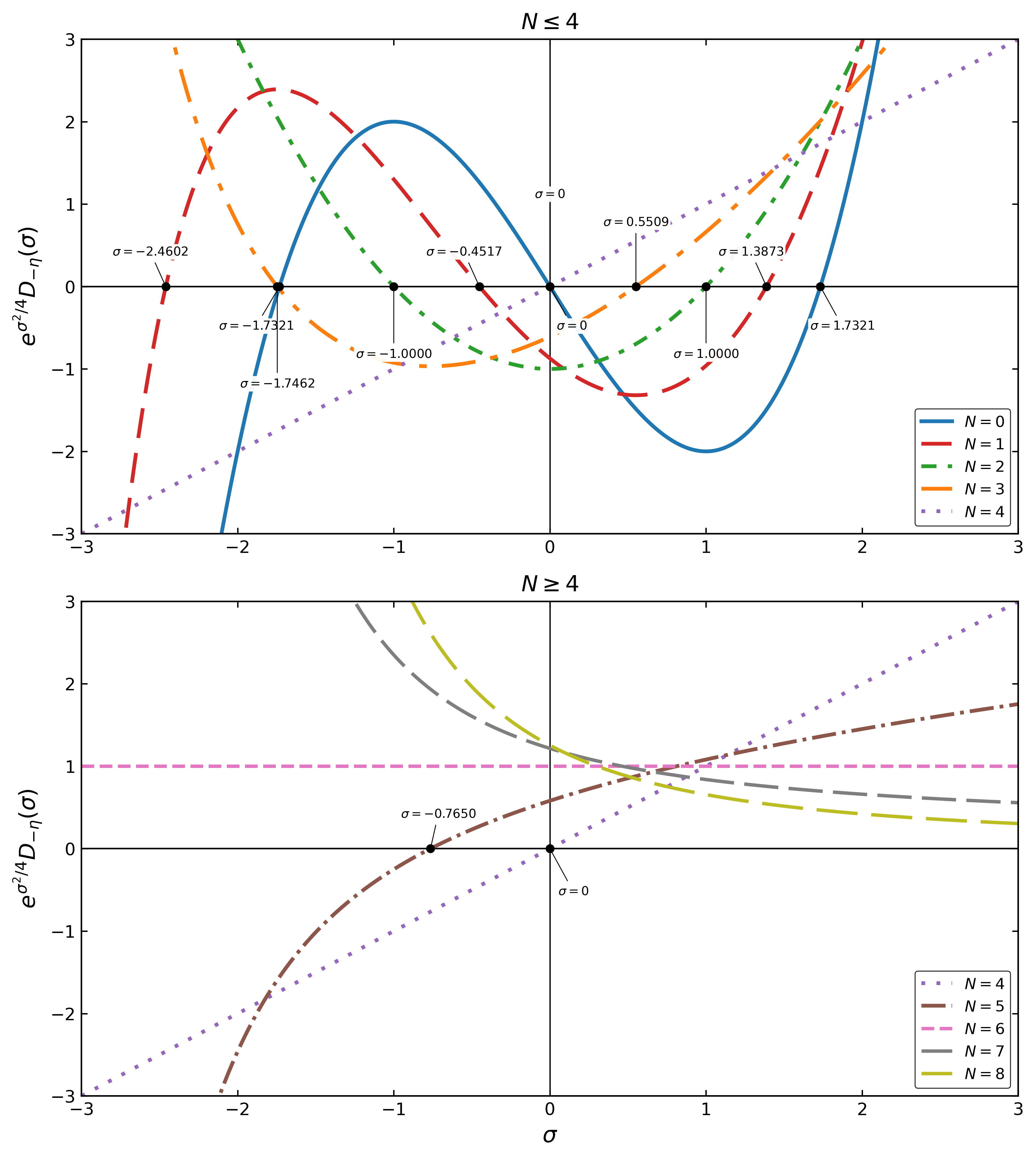}
\caption{Real-axis denominators \(e^{\sigma^2/4}D_{-\eta}(\sigma)\). Real zeros are marked by black points. \(\cN=4\) has only the zero \(\sigma=0\). The upper panel shows the positive-real zeros present for \(\cN<4\). The lower panel shows that no positive-real zero remains for \(\cN>4\).}
\label{fig:denominatorcurves}
\end{figure}

\cref{fig:denominatorcurves} shows the motion of the rightmost real zero. For \(\cN<4\), positive-real zeros are present. At \(\cN=4\), only the zero at \(\sigma=0\) remains. For \(\cN>4\), no positive-real zero survives. This is the transition between pole-driven Regge growth and the absence of such positive-real Mellin poles.

\subsection*{Special theories and the low-supersymmetry regime}
\label{subsec:specialtheories}

The closed forms in this subsection are not new. Their role is to show how the known special theories fit naturally into the rank-two picture.

Since \(R(\sigma)\) was defined with an explicit prefactor of \(\eta\), the variable \(R\) becomes degenerate at \(\eta=0\) and no longer captures the nontrivial \(\cN=6\) solution. It is therefore more natural to divide out that prefactor and introduce
\begin{equation}
 Y(\sigma)=\lim_{\eta\to 0}\frac{R(\sigma)}{\eta}
 =\sqrt{-\alpha t}\,\frac{f_\omega^{(6)}}{\omega}.
 \label{eq:Ydef}
\end{equation}
This is the smooth limit of the neighboring-period ratio at the rung \(\widehat J_6(\sigma)=1\). Dividing \cref{eq:riccati} by \(\eta\) and taking \(\eta\to 0\) gives
\begin{equation}
 Y'(\sigma)=\sigma Y(\sigma)-1.
 \label{eq:Yeq}
\end{equation}
Multiplying by \(e^{-\sigma^2/2}\) and integrating from \(\sigma\) to \(\infty\) yields
\begin{equation}
 Y(\sigma)
 =e^{\sigma^2/2}\int_\sigma^\infty e^{-u^2/2}\,\dd u
 =\sqrt{\frac{\pi}{2}}\,e^{\sigma^2/2}\,
 \erfc\!\left(\frac{\sigma}{\sqrt2}\right).
 \label{eq:Ysolution}
\end{equation}
At the level of the full resummed factor, \cref{eq:N6doublefact} gives
\begin{equation}
 M_{4,\mathrm{DL}}^{(6)}(s,t)
 =
 \sum_{m=0}^{\infty}\frac{(2m-1)!!}{(2m)!}g^m
 =
 e^{g/2}.
 \label{eq:N6exponential}
\end{equation}
So \(\cN=6\) is exactly solvable in elementary terms.

The theory \(\cN=4\) is the critical one. Its exact cancellation was already obtained from the discrete recursion. The same result follows from the special-function identity \(D_1(\sigma)=\sigma e^{-\sigma^2/4}\), which gives
\begin{equation}
 R(\sigma)=-\frac{1}{\sigma},
 \qquad
 \frac{f_\omega^{(4)}}{\omega}=\frac{1}{\omega},
\end{equation}
hence \(f_\omega^{(4)}\equiv 1\).

For \(\cN<4\), the dominant Regge behavior is controlled by the rightmost positive-real Mellin pole. Since \(\eta<-1\), the denominator develops a zero at some \(\sigma_*>0\). By \cref{eq:universalresidue}, the corresponding pole \(\omega_*=\sigma_*\sqrt{-\alpha t}\) has residue \(-1/\eta>0\). Therefore
\begin{equation}
 M_{4,\mathrm{DL}}^{(\cN)}(s,t)
 \sim
 -\frac{1}{\eta}
 \left(\frac{s}{-t}\right)^{\sigma_*\sqrt{-\alpha t}},
 \qquad
 \frac{s}{-t}\to\infty,
 \qquad
 \cN<4.
 \label{eq:Nlt4asymptotic}
\end{equation}
The first four low-supersymmetry theories are listed in \cref{tab:lowNzeros}.

\begin{table}[t]
\centering
\small
\renewcommand{\arraystretch}{1.15}
\begin{tabular}{@{}cccc@{}}
\toprule
\(\cN\) & \(\eta\) & rightmost positive zero \(\sigma_*\) & residue \(-1/\eta\) \\
\midrule
\(0\) & \(-3\)   & \(\sqrt{3}\)           & \(1/3\) \\
\(1\) & \(-5/2\) & \(1.3872729596\ldots\) & \(2/5\) \\
\(2\) & \(-2\)   & \(1\)                  & \(1/2\) \\
\(3\) & \(-3/2\) & \(0.5508550482\ldots\) & \(2/3\) \\
\bottomrule
\end{tabular}
\caption{Rightmost positive zeros of the denominator and the corresponding universal residues in the pole-dominated regime \(\cN<4\).}
\label{tab:lowNzeros}
\end{table}

As \(\cN\) increases from \(0\) to \(4\), the rightmost positive Mellin pole moves left and disappears at the critical theory.

The even theories below \(\cN=6\) can also be written in closed form. For \(\cN=2\), one has \(\eta=-2\), hence
\begin{equation}
 D_2(\sigma)=(\sigma^2-1)e^{-\sigma^2/4},
 \qquad
 D_1(\sigma)=\sigma e^{-\sigma^2/4},
\end{equation}
so
\begin{equation}
 \frac{f_\omega^{(2)}}{\omega}
 =\frac{\omega}{\omega^2-b}
 =\frac12\left(\frac{1}{\omega-\sqrt b}+\frac{1}{\omega+\sqrt b}\right).
\end{equation}
Mellin inversion gives
\begin{equation}
 M_{4,\mathrm{DL}}^{(2)}(s,t)
 =\cosh\!\left(\sqrt{-\alpha t}\,\ln\!\frac{s}{-t}\right).
 \label{eq:N2exact}
\end{equation}
For pure Einstein gravity, \(\cN=0\) and \(\eta=-3\), so
\begin{equation}
 \frac{f_\omega^{(0)}}{\omega}
 =\frac{\omega^2-b}{\omega(\omega^2-3b)}
 =\frac{1}{3\omega}
 +\frac{1}{3}\left(
 \frac{1}{\omega-\sqrt{3b}}
 +\frac{1}{\omega+\sqrt{3b}}
 \right),
\end{equation}
hence
\begin{equation}
 M_{4,\mathrm{DL}}^{(0)}(s,t)
 =\frac{1}{3}
 +\frac{2}{3}\cosh\!\left(\sqrt{-3\alpha t}\,\ln\!\frac{s}{-t}\right).
 \label{eq:N0exact}
\end{equation}
These are the first two nontrivial outputs of the even recursion. For \(\cN=3\) and \(\cN=1\), the answer remains the same period ratio, but the rightmost positive-real zero still controls the leading Regge growth.

\section{Twisted intersection derivation of the reduction identity}
\label{sec:intersection}

The reduction identity \cref{eq:I2reduction} already closes the differential system. The purpose of this section is to recover the same identity from a second method that is systematic and potentially useful beyond the present one-variable example. The idea is simple. Instead of reducing the third moment by a direct total derivative, one studies a pairing between weighted differential forms. For related intersection-theoretic treatments of regulated Gaussian integrals in quantum mechanics and quantum field theory, see \cite{CacciatoriMastrolia2022}. The final answer is the same reduction, but the route to it is algorithmic.

\subsection*{A short dictionary}

We need only a small amount of terminology. For the one-variable problem on $X=\mathbb C\setminus\{0\}$, 
it is useful to recall the twist already introduced in \cref{eq:twist}. In this section, \(\omega=\dd\ln u\) denotes the logarithmic one-form associated with the twist, not the Mellin variable of \cref{eq:mellinrep}. Explicitly,
\begin{equation}
 \omega=\dd\ln u=
 \left(\frac{\eta-1}{z}-z-\sigma\right)\dd z.
 \label{eq:omegadef}
\end{equation}
For generic values of \((\sigma,\eta)\), the critical points of \(\ln u\) are the solutions of $z^2+\sigma z-(\eta-1)=0$, 
namely
\begin{equation}
 z_\pm=\frac{-\sigma\pm\sqrt{\sigma^2+4(\eta-1)}}{2}.
\end{equation}
Thus, away from the discriminant \(\sigma^2+4(\eta-1)=0\), the relevant twisted cohomology and homology are generically two-dimensional, \(\dim H_\omega^1(X)=\dim H_1^\omega(X)=2\). This matches the existence of two master periods, equivalently the fact that the scalar differential equation is second order and the associated first-order system has rank two \cite{CacciatoriMastrolia2022,LeePomeransky2013}.
The ordinary exterior derivative is then replaced by the weighted derivative
\begin{equation}
 \nabla_\omega=\dd+\omega\wedge.
\end{equation}
A weighted differential form is considered only up to terms of the form \(\nabla_\omega\chi\). The goal is to decompose one such form in terms of a basis of simpler ones. The coefficients of that decomposition are obtained from a pairing called the intersection pairing. In one complex variable, this pairing reduces to a residue computation. This is why the present example can be worked out completely by hand.

In the notation of \cite{MastroliaMizera2019}, if \(C_{ij}=\langle\varphi_i\mid\varphi_j\rangle\) is the intersection matrix of a basis \(\{\varphi_i\}\), and if \(v_j=\langle\varphi\mid\varphi_j\rangle\) is the projection of a target form \(\varphi\), then the coefficients of the reduction are obtained by solving the linear system \(C^Tc=v\). For the graviton problem this becomes a \(2\times 2\) exercise.

For the holomorphic one-forms used below, being \(\nabla_\omega\)-closed is automatic. If \(\varphi=f(z)\,\dd z\) with \(f\) holomorphic on \(X\), then \(\dd\varphi=0\) in one variable and \(\omega\wedge\varphi=0\) because two one-forms wedge to zero. Twisted cohomology in degree one is therefore the space of holomorphic one-forms modulo weighted exact forms.

Choose the basis
\begin{equation}
 \varphi_0=\dd z,
 \qquad
 \varphi_1=z\,\dd z.
 \label{eq:basiscocycles}
\end{equation}
The same representatives will be used for the dual basis in \(H^1_{-\omega}(X)\). The intersection pairing is then computed from local solutions of
\begin{equation}
 \nabla_\omega\Psi_p=\varphi,
 \qquad
 \partial_z\Psi_p+\left(\frac{\eta-1}{z}-z-\sigma\right)\Psi_p=f(z).
 \label{eq:localeq}
\end{equation}
If \(\psi=g(z)\,\dd z\), the pairing is
\begin{equation}
 \langle \varphi\mid\psi\rangle
 =\sum_p\Res_{z=p}(\Psi_p\,\psi).
 \label{eq:pairingdef}
\end{equation}
In our basis, the contribution from \(z=0\) vanishes for generic \(\eta\). Everything reduces to Laurent expansions at infinity.

\subsection*{Local solutions and the reduction coefficients}

For \(\varphi_0=\dd z\), write
\begin{equation}
 \Psi_0(z)=\sum_{n\ge 0}a_n z^{-n-1}.
 \label{eq:Psi0ansatz}
\end{equation}
Then \(\partial_z\Psi_0(z)=\sum_{n\ge 0}-(n+1)a_n z^{-n-2}\), and
\begin{equation}
 \left(\frac{\eta-1}{z}-z-\sigma\right)\Psi_0(z)
 =
 \sum_{n\ge 0}(\eta-1)a_n z^{-n-2}
 -\sum_{n\ge 0}a_n z^{-n}
 -\sigma\sum_{n\ge 0}a_n z^{-n-1}.
\end{equation}
Matching coefficients in the local equation gives \(a_0=-1\) and \(a_1=\sigma\), so
\begin{equation}
 \Psi_0(z)=-z^{-1}+\sigma z^{-2}+\mathcal O(z^{-3}).
 \label{eq:Psi0series}
\end{equation}

For \(\varphi_1=z\,\dd z\), write
\begin{equation}
 \Psi_1(z)=\sum_{n\ge 0}b_n z^{-n}.
\end{equation}
The same matching gives \(b_0=-1\), \(b_1=\sigma\), and \(b_2=-(\sigma^2+\eta-1)\), and therefore
\begin{equation}
 \Psi_1(z)
 =-1+\sigma z^{-1}-(\sigma^2+\eta-1)z^{-2}+\mathcal O(z^{-3}).
 \label{eq:Psi1series}
\end{equation}

For the target form \(\varphi=z^2\,\dd z\), write
\begin{equation}
 \Psi_2(z)=\sum_{n\ge 0}d_n z^{1-n}.
\end{equation}
Matching coefficients gives \(d_0=-1\), \(d_1=\sigma\), \(d_2=-(\sigma^2+\eta)\), and \(d_3=\sigma^3+(2\eta-1)\sigma\). Hence
\begin{equation}
 \Psi_2(z)
 =-z+\sigma-(\sigma^2+\eta)z^{-1}
 +\bigl(\sigma^3+(2\eta-1)\sigma\bigr)z^{-2}
 +\mathcal O(z^{-3}).
 \label{eq:Psi2series}
\end{equation}

We now compute the intersection matrix. Using
\begin{equation}
 \Res_{z=\infty}(h(z)\,\dd z)=-[z^{-1}]\,h(z),
 \label{eq:resinfty}
\end{equation}
one finds from \cref{eq:Psi0series,eq:Psi1series}
\begin{equation}
 C_{00}=1,
 \qquad
 C_{01}=-\sigma,
 \qquad
 C_{10}=-\sigma,
 \qquad
 C_{11}=\sigma^2+\eta-1.
\end{equation}
Therefore
\begin{equation}
 C=
 \begin{pmatrix}
 1 & -\sigma\\
 -\sigma & \sigma^2+\eta-1
 \end{pmatrix},
 \qquad
 \det C=\eta-1.
 \label{eq:Cmatrix}
\end{equation}
The special value \(\eta=1\) makes this particular basis degenerate, but it does not signal a collapse of the underlying rank-two twisted cohomology. It only means that this basis is ill-adapted at that point and should be replaced by a nondegenerate one. From \cref{eq:Psi2series}, the projection vector of \(\varphi=z^2\,\dd z\) is
\begin{equation}
 v_0=\langle \varphi\mid\varphi_0\rangle
 =-\,[z^{-1}]\Psi_2(z)
 =\sigma^2+\eta,
\end{equation}
\begin{equation}
 v_1=\langle \varphi\mid\varphi_1\rangle
 =-\,[z^{-1}](z\Psi_2(z))
 =-\sigma^3-(2\eta-1)\sigma.
 \label{eq:vvector}
\end{equation}
Solving \(C^Tc=v\) then gives
\begin{equation}
 C^{-1}
 =\frac{1}{\eta-1}
 \begin{pmatrix}
 \sigma^2+\eta-1 & \sigma\\
 \sigma & 1
 \end{pmatrix},
\end{equation}
and
\begin{align}
 c
 &=
 \frac{1}{\eta-1}
 \begin{pmatrix}
 \sigma^2+\eta-1 & \sigma\\
 \sigma & 1
 \end{pmatrix}
 \begin{pmatrix}
 \sigma^2+\eta\\
 -\sigma^3-(2\eta-1)\sigma
 \end{pmatrix}
 =
 \begin{pmatrix}
 \eta\\
 -\sigma
 \end{pmatrix}.
 \label{eq:csolution}
\end{align}
Therefore
\begin{equation}
 z^2\,\dd z
 \equiv
 \eta\,\dd z-\sigma z\,\dd z
 \qquad \text{in } H^1_\omega(X).
 \label{eq:cohomologicalreduction}
\end{equation}
Multiplying by the weight \(u\) and integrating along the positive real axis gives back exactly \cref{eq:I2reduction}. The direct total-derivative argument and the intersection calculation therefore lead to the same reduction. The value of the second method is that it can plausibly be turned into a systematic procedure for more complicated sectors of the all-orders four-graviton scattering amplitude.

\section{Conclusions and Outlook}

The double-logarithmic sector of four-graviton scattering in \(\cN\)-extended supergravity was solved long ago in terms of parabolic-cylinder functions \cite{BartelsLipatovSabioVera2014}. The point of the present work is not to replace that solution, but to show that it has a simpler organization than is apparent in its original form. The basic observation is that the Mellin partial wave is controlled by two neighboring normalized periods, introduced in \cref{eq:Jhatdef}. Their ratio gives the physical Mellin partial wave directly, as in \cref{eq:fJhatratio}. In this sense the problem is genuinely rank two. Two functions are enough to reconstruct the full answer. These two periods satisfy a first-order differential system in the Mellin variable \(\sigma\), namely \cref{eq:PfaffianJhat}, and they also satisfy a discrete recursion in the theory label \(\cN\), namely \cref{eq:Jhatrecurrence}. The first relation tells us how the two periods vary with \(\sigma\). The second moves from one supergravity theory to the next. Taken together, they reorganize the known Weber solution into a single framework that treats the whole supergravity family uniformly.

The comparison between the positive-ray integral and the contour formula also becomes precise in this language. The positive real axis gives the simplest convergent representative whenever \(\Re(\eta)>0\). The contour formula, through \cref{eq:contourEulerRelationExpanded,eq:JhatContour}, keeps track of the monodromy around the branch point and continues the same object beyond that naive convergence region. In this way the old contour representation and the present period language are seen to describe the same analytically continued solution.

The same one-step reduction can also be recovered by intersection theory. The cohomological reduction formula \cref{eq:cohomologicalreduction} reproduces the same relation obtained by direct integration by parts. This gives an independent check of the construction and shows that the reduction is not accidental. It is built into the geometry of the weighted integrals themselves.

Several physical features become especially transparent in this form. The exact disappearance of the double-logarithmic sector at \(\cN=4\) is built into the discrete recursion, see \cref{eq:N4exactcancellation}. The theory with \(\cN=6\) is the natural starting point of the even family, as seen in \cref{eq:J6base}. The lower even theories are then generated by Hermite polynomials, as in \cref{eq:Heidentification,eq:evenHermitePartialFraction}, which immediately yields the closed forms for \(\cN=2\) and \(\cN=0\) in \cref{eq:N2exact,eq:N0exact}. The Mellin residues are universal, see \cref{eq:universalresidue}, and the change at \(\cN=4\) cleanly separates theories with positive-real Mellin poles from those without them, see \cref{eq:Nthresholdclassification}.

The all-order formulas are also consistent with the available loop data. In maximal supersymmetry the logarithm of the resummed factor reproduces the expected expansion in \cref{eq:N8logseries}, and the leading two-loop and three-loop Regge logarithms are matched by \cref{eq:F2leadinglog,eq:HM3loopleadingours}. The period formulation therefore reorganizes the known solution without changing its perturbative content.

We have not solved the full Regge limit of gravity. The more modest claim is that one closed infrared-finite subsector admits a controlled description in terms of two weighted integrals, one first-order differential system, one discrete recursion in \(\cN\), and one independent cohomological derivation. This is complementary to the main amplitude-based approaches to high-energy gravity. Double-copy methods explain how supergravity amplitudes are assembled from gauge-theory data at fixed loop order \cite{BoucherVeronneauDixon2011}. Eikonal and Wilson-line methods describe graviton reggeization and exponentiation directly in impact-parameter space and in effective high-energy descriptions \cite{DiVecchiaHeissenbergRussoVeneziano2024}. The present paper isolates the double-logarithmic sector as a fully worked example where those exact amplitudes can also be described through periods, differential equations, and intersection numbers.

Several future directions suggest themselves. If the leading double-logarithmic sector closes on two functions, then subleading logarithmic sectors may require larger systems. If the relevant weighted integrals depend on more than one variable, then one is naturally led to multivariable period problems. The intersection-based derivation presented here suggests that such cases may still be tractable with modern reduction methods. It is also natural to ask whether the gauge-theory sectors that enter the double-copy story admit a parallel description. Finally, the explicit formulas in \cref{eq:N0exact,eq:N2exact} provide concrete testing grounds for future all-order and fixed-loop analyses.

\section*{Acknowledgments}

The author thanks Pierpaolo Mastrolia for a careful reading of the manuscript and  acknowledges support from the Spanish Agencia Estatal de Investigaci\'on through grants PID2022-142545NB-C22 and IFT Centro de Excelencia Severo Ochoa CEX2020-001007-S, funded by MCIN/AEI/10.13039/501100011033 and by ERDF ``A way of making Europe'', and from the European Union's Horizon 2020 research and innovation programme under grant agreement No.~824093.

\end{document}